\documentclass[twocolumn,prl,showpacs,preprintnumbers,amsmath,amssymb]{revtex4}

\usepackage{graphicx}
\usepackage{dcolumn}
\usepackage{bm}

\begin{document}


\title{Knudsen-to-Hydrodynamic Crossover in Liquid $^3$He in a High-Porosity Aerogel}

\author{H. Takeuchi}
\author{S. Higashitani}
\author{K. Nagai}
\affiliation{ Graduate School of Integrated Arts and Sciences, Hiroshima
University, Kagamiyama 1-7-1, Higashi-Hiroshima 739-8521, Japan }
\author{H. C. Choi}
\author{B. H. Moon}
\author{N. Masuhara}
\author{M. W. Meisel}
\author{Y. Lee}
\affiliation{National High Magnetic Field Laboratory and Department of Physics, University of Florida, Gainesville, FL 32611-8440, USA}%
\author{N. Mulders}
\affiliation{Department of Physics and Astronomy, University of
Delaware, Newark, DE 19716, USA}

\date{\today}

\begin{abstract} 
 We present a combined experimental and theoretical study of the drag force
 acting on a high porosity aerogel immersed in liquid ${}^3$He and its
 effect on sound propagation.  The drag force is characterized by the
 Knudsen number, which is defined as the ratio of the quasiparticle mean
 free path to the radius of an aerogel strand.  Evidence of the
 Knudsen-hydrodynamic crossover is clearly demonstrated by a drastic change in the
 temperature dependence of ultrasound attenuation in 98~\% porosity aerogel. Our theoretical analysis shows  that the frictional sound damping caused by the drag force is governed by distinct laws in the
 two regimes, providing excellent agreement with the experimental observation.
\end{abstract}


\maketitle

The understanding of complex flow phenomena, such as in porous media, is a challenging
goal of fluid physics.  The pioneering work by Darcy \cite{Darcy1856}
established empirically a linear relation between the fluid flux of
laminar viscous flow in porous media and the total pressure drop in balance
with viscous drag force. The Darcy law provides a constitutive relation in
hydrodynamics and has been applied to various kinds of flow phenomena such
as water flow in aquifers, blood flow in vessels, and petroleum flow
through sandstone and gravel \cite{Collins1961, Molz1989, Nagy2008}.  In
general, however, the hydrodynamic description breaks down when the mean
free path (mfp) of fluid particles exceeds substantially the characteristic
dimension of the porous material. In this case, the fluid mass is carried by
the Knudsen diffusion \cite{Knudsen}, {\it i.e.\,}, successive ballistic flights of
individual particles followed by scattering at walls. The Knudsen diffusion,
as well as the viscous flow, is of particular interest for gas transport
in nanofabricated systems \cite{K_transport1,K_transport2}, which have
recently attracted significant attention in the context of nanofluidics
\cite{ref_nanofluidics}. The investigation of the mfp dependent flow regimes
is also of fundamental importance towards a unified understanding of the
transport properties in complex porous materials existing in nature.

Due to the unique porous structure of aerogel and the fascinating low temperature properties of liquid $^3$He, the $^3$He-aerogel system presents an opportunity that allows investigations of a wide range of physical phenomena. The superfluid phase of $^3$He in highly porous aerogel is an analog of unconventional
superconductors with pair-breaking impurities \cite{Porto1995,Sprague,NomuraGHLMH2000PRL85,ChoiPRL2007,Gervais,MoonPRB2010} and has been the subject of intense study since the first observations \cite{Porto1995,Sprague} of superfluid transitions in this system. In the normal liquid, one can study the Fermi liquid behavior ranging from transport properties \cite{NomuraGHLMH2000PRL85,LeeJLTP,Ichikawa2001,HigashitaniMIYN2002PRL89,ReeJLTP2002,SauPRB2005} to diverse flow phenomena \cite{EinPRL} in the presence of impurity-like disorder with a topologically antithetical geometry to the other conventional porous media.  In this Letter, we present the first experimental and theoretical study on the Knudsen-hydrodynamic crossover in the normal liquid $^3$He confined in 98\% porosity silica aerogel. 

\begin{figure}[t]
 \centering
 \includegraphics[width=1. \linewidth]{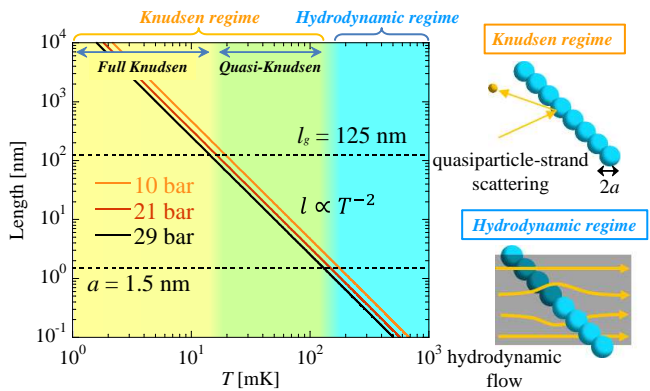}
 \caption{Characteristic length scales of 98\% porosity aerogel and the temperature dependence of the 
 quasiparticle mean free path for three pressures, $l \propto 1/T^2$ in pure liquid $^3$He. The mfp was
 extracted from the experimental data of viscosity \cite{Wheatley1975RMP47}.
 The whole temperature range is divided into three transport regimes classified by the hydrodynamic, quasi-Knudsen, and full Knudsen regime. For $T \lesssim 15$~mK (full Knudsen regime) the quasiparticle mfp is limited by the geometrical mfp $l_g$.  Above $\sim 15$~mK, the mutual
 quasiparticle collision becomes dominant, but the drag force is still in the Knudsen regime
 as far as $l > a$ (quasi-Kundusen regime).  Around 150~mK (hydrodynamic regime), where $l \sim a$,
 the Knudsen-hydrodynamic crossover of the drag force occurs.  The figures on the right
 show a schematic illustration of two extreme flow regimes.
}
 \label{fig:length}
\end{figure}%

Liquid $^3$He is a paradigm of Fermi liquids in which the transport property
is governed by low energy thermal excitations called quasiparticles. Since
the collision processes between the quasiparticles are restricted by the
Pauli exclusion principle, the quasiparticle mfp, $l$, can be much longer than
the interparticle distance. The mfp increases with decreasing
temperature $T$, following a $1/T^2$-dependence \cite{Landau_Lifshitz}, and approaches
$\sim 10$ $\mu$m near the superfluid transition temperature, $T_c \simeq 2$~mK (see Fig.\ \ref{fig:length}).  On the other hand, aerogel,
an interconnected network of thin strands of aggregated silica
particles, introduces additional length scales: the strand diameter $2a$, the inter-strand distance $\zeta$, and the geometrical mean free path $\l_{g}$.  Specifically, in 98\% porosity aerogel, $2a \simeq 3$~nm, $\zeta \approx 25-40$~nm, and finally $\l_{g} \approx 100-150$~nm.  There is no well-defined pore structure in aerogel in contrast to the nanochannels used in the
previous works \cite{K_transport1,K_transport2} on the Knudsen-hydrodynamic crossover.

To understand the effect of the high porosity aerogel on the mass transport
in liquid $^3$He, we performed extensive ultrasound transmission
measurements over a wide temperature range up to 200 mK.  Based on the
experimental results, we shall show that the transport properties in aerogel
are characterized by two length scales of aerogel, the strand radius $a$ and
the geometrical mfp $l_g$, separating three transport regimes as summarized
in Fig.\ \ref{fig:length}.

The acoustic cavity used in this experiment is formed by two longitudinal LiNbO$_3$ transducers
used as a transmitter and a receiver. The transducers are separated by a Macor spacer maintaining a $3.05 \pm 0.02$~mm parallel gap. The pair of transducers were carefully selected to achieve the best match in the fundamental frequency (9.5~MHz) and the frequency response. The 98\% porosity aerogel sample 
was grown between the transducers {\it in situ} to prevent any irregular
reflections at the transducer-aerogel boundaries. Temperature was determined by a melting pressure
thermometer for $T >1 $~mK and a Pt NMR thermometer for $T < 1$~mK.  The pressure of the cell was monitored throughout the measurements by a capacitive pressure sensor incorporated as a part of the experimental cell.  A commercial spectrometer, LIBRA/NMRKIT II (Tecmag Inc., Houston, TX),
was used to excite the transmitter and observe the propagated sound signal
from the receiver. A 1~$\mu$s pulse was sent to the
transmitter and the signal from the receiver was amplified with a preamplifier (Miteq AU1114)
before being transferred to the spectrometer. A detailed description of the experimental setup can be
found elsewhere \cite{LeeJLTP}.

\begin{figure}[t]
 \centering
 \includegraphics[width=.9 \linewidth]{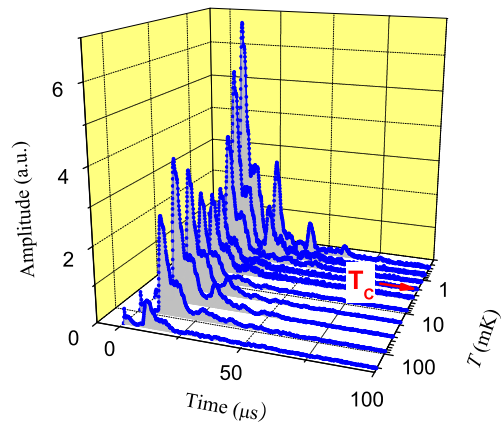}
 \caption{Receiver signal outputs at 29 bar for selected temperatures. The superfluid transition in aerogel is demarcated by an arrow.}
 \label{fig:experiment}
\end{figure}

In Fig.\ \ref{fig:experiment}, the receiver signal outputs acquired at 29 bar are plotted
for selected temperatures. The signals for all temperatures rise around 8~$\mu$s. Below 0.5~mK, in the low attenuation regime, one can see the three echoes following the initial peak.  
From the time-of-flight measurements, the sound velocity $c$ was obtained to
be $c=330$~m/s at 29~bar. The temperature dependence of the sound velocity is
less than our measurement error for the whole temperature range presented in this work.

\begin{figure}[tb]
 \centering
 \includegraphics[width=.8 \linewidth]{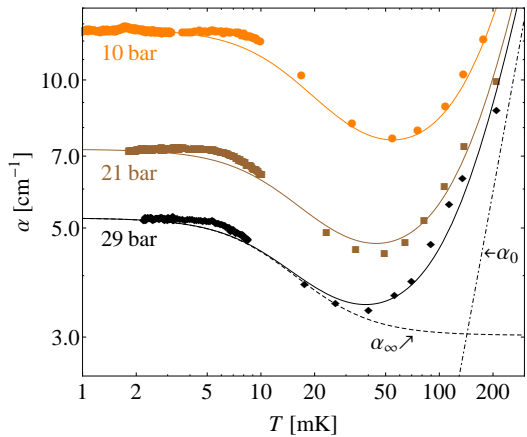}
 \caption{Ultrasound attenuation ($\omega/2\pi = 9.5$ MHz) in the normal fluid above the aerogel superfluid transition. The solid symbols represent the experimental data at
three different pressures.  The solid lines are the results of fit to the data using
 Eqs.\ \eqref{alpha}-\eqref{alphaf} with the Cunningham constant $A =
 6/5$. The fitting parameters are $(l_{\rm tr},l_a)=(115.4, 157.2)$~nm for
 10 bar, $(108.7,147.1)$~nm for 21 bar, and $(100.0, 135.4)$~nm for 29 bar.  The
 dashed lines are the sound attenuation calculated in the Knudsen ($\alpha_\infty$) and hydrodynamic  ($\alpha_0$) limit (see text).}  \label{fig:alpha}
\end{figure}

The relative sound attenuation, $\alpha$, was determined using two different
methods.  One was to use the peak value of the initial signal and the other
was to use the integrated signal of the initial peak from the rising edge to
23~$\mu$s. Both methods produced consistent results within 1\% in the low
attenuation regime and within 5 \% in the high attenuation regime. The
relative attenuation was then converted into the absolute attenuation using
a reference value at 0.4 mK where 3 echoes were clearly observed
 owing to the low attenuation in the superuid phase. The peaks of the echoes exhibit an exponential
decay, enabling us to determine accurately the absolute attenuation.
 In Fig.\
\ref{fig:alpha}, the attenuation data obtained by the second method are
plotted as a function of temperature $T$. The data at three different
pressures are shown above the superfluid transition in aerogel.  The
attenuation $\alpha$ is almost independent of $T$ near $T_c$ but decreases
gradually with increasing $T$.  This behavior is in good agreement with
those observed in previous ultrasound attenuation measurements
\cite{NomuraGHLMH2000PRL85}. However, our higher temperature measurements
revealed a new behavior which exhibits a minimum around 50~mK followed by a
rather rapid increase in attenuation with temperature.

As the first step of analysis of the experimental results, let us estimate
$l_g$ in 98 \% aerogel. The simplest estimate can be made by modeling the
aerogel as a dilute random distribution of spheres with radius $a$ (sphere
model). Noting that the sphere density $n_a$ is related to the aerogel
porosity $\phi$ by $(4/3)\pi a ^3n_a = 1-\phi$, we obtain $l_g = 1/\pi a^2
n_a = (4/3)a/(1-\phi)$. Substitution of $a = 1.5$~nm and $\phi = 0.98$
yields $l_g = 100$~nm. A more realistic model of randomly oriented cylinders
\cite{Thuneberg1998} also leads to a similar value $l_g = 2a/(1-\phi) =
150$~nm.  These estimations are consistent with those from an empirical
formula extracted from computer simulations based on a diffusion limited
cluster aggregation model \cite{PorPRB1999}.  In Fig.\ \ref{fig:length},
$l_g$ is compared with $l \propto 1/T^2$ in pure liquid $^3$He. We observe
$l_g \approx l$ at around 15~mK.

At low temperatures where $l > l_g$, the transport properties are governed
by the geometry of aerogel structure.
This point is clearly demonstrated by regarding
the quasiparticle-quasiparticle and quasiparticle-aerogel
collisions as two independent scattering sources. Then,
for example,
 the effective viscosity coefficient
in the $^3$He-aerogel system has the form
\begin{equation}
 \eta_{\rm eff} = \frac{1}{5}{np_F}(1/l_a + 1/l)^{-1}
  = \frac{1}{5}\frac{np_Fl_a}{1 + T^2/T_a^2}
  \label{eta}
\end{equation}
with $n$ being the $^3$He number density, $p_F$ the Fermi momentum, $l_a$ a
temperature independent mfp of the order of $l_g$, and $T_a$ the temperature
at which $l = l_a$.  At high temperatures, $T \gg T_a$, Eq.\ \eqref{eta} is
reduced to the conventional gas-kinetic formula for a pure Fermi liquid. As
temperature decreases, the quasiparticle scattering processes become
gradually dominated by the aerogel-quasiparticle collisions, so that
$\eta_{\rm eff}$ saturates to the geometry-limited value of
$(1/5)np_{F}l_{a}$. Similar saturation behavior is observed for the
spin-diffusion coefficient of $^3$He in 98 \% aerogel \cite{SauPRB2005}.

Now we discuss the sound propagation in aerogel.  The sound frequency $\omega =
2\pi\times 9.5$ MHz used in our measurements should be compared to
the effective quasiparticle scattering rate, $1/\tau_{\rm eff} = v_F(1/l_a +
1/l)$. The Fermi velocity $v_F$ in liquid $^3$He is $\simeq 50$~m/s
\cite{HalperinV1990}. Since $1/\tau_{\rm eff} > v_F/l_a \agt 10^8$
s$^{-1}$, our ultrasound measurements are in the collision-dominated regime,
$\omega < 1/\tau_{\rm eff}$.  Therefore, the sound propagation can then be described by
the Navier-Stokes equation
\begin{align}
 \rho \partial_t \bm{v} = - \nabla P + \eta_{\rm eff} (\Delta \bm{v} +
 \frac{1}{3}\nabla{\rm div}\bm{v}) - \bm{F},
\end{align}
where $\rho=mn$ is the $^3$He mass density, $\bm{v}$ the velocity field of
liquid $^3$He, $P$ pressure, and $\bm{F}$ the drag force (per unit volume)
exerted on aerogel.  This drag force gives rise to the coupled oscillations of liquid $^3$He and
aerogel \cite{Ichikawa2001, HigashitaniMIYN2002PRL89}.  Since the sound wavelength of interest is much longer than the
distance between aerogel strands, the aerogel oscillations can be described
within a continuum approximation,
\begin{align}
 \rho_a \partial_t^2 \bm{u}_a = \rho_a c_a^2 \Delta \bm{u}_a + \bm{F},
\end{align}
where $\bm{u}_a$, $\rho_a \simeq 0.044$ g/cm${}^3$, and $c_a \simeq 50$ m/s
are the displacement vector, the mass density, and the speed of longitudinal sound in
aerogel itself, respectively.


To proceed further, we need the constitutive relation for $\bm{F}$. Let Kn
be the Knudsen number defined by Kn $= l/a \propto 1/T^2$.  Since the
smallest length scale of aerogel is the strand radius $a$, the hydrodynamic
description of the drag force is justified for Kn $< 1$. In the hydrodynamic
limit, Kn $\to 0$, the drag force density obeys the Darcy law
\begin{eqnarray}
 {\bm F}_0 = \frac{\eta}{\kappa}\bm{v}_{\rm rel}.
  \label{eq:DarcyLaw}
\end{eqnarray}
Here $\bm{v}_{\rm rel} = \bm{v} - \partial_t \bm{u}_a$ is the relative
velocity field of liquid $^3$He and aerogel, $\eta = (1/5)np_Fl$ the
viscosity of pure liquid $^3$He, and $\kappa$ the permeability.  In the
sphere model for aerogel, the Darcy law is equivalent to the Stokes law
${\bm F}_{0} / n_a = 6\pi a \eta {\bm v}_{\rm rel}$, and then the
permeability takes the form, $\kappa={al_g}/{6}$.

In the opposite limit, Kn $ \to \infty$, where the Darcy law breaks down,
the drag force is determined by the collisions between aerogel and
individual quasiparticles (see Fig.\ \ref{fig:length}). Ichikawa {\it et
al.}\ \cite{Ichikawa2001} discussed the drag force on aerogel by modeling it
as randomly distributed impurities. By calculating the momentum transfer
from liquid $^3$He to aerogel using the impurity-quasiparticle collision
integral, they obtained the following expression for the drag force density:
\begin{eqnarray}
 {\bm F}_{\infty}=\frac{np_F}{l_{\rm tr}}\bm{v}_{\rm rel},
  \label{eq:KnudsenLimit}
\end{eqnarray}
where $l_{\rm tr}$ is the transport mfp. Equation \eqref{eq:KnudsenLimit}
has an anticipated form of the drag force density in the Knudsen limit, Kn $\to \infty$. 
Specifically, consider a situation in which an object with a cross
section $\sigma$ is moving with a constant velocity $-\bm{v}_{\rm rel}$ in
``gas'' of quasiparticles with momentum $p_F$. The body would experience collisions with
the quasiparticles of number $n\sigma|\bm{v}_{\rm rel}|$ in unit time and experiences a force
$\sim p_Fn\sigma\bm{v}_{\rm rel}$. Dividing it by the total volume $V$ and
replacing $\sigma/V$ by $1/l_{\rm tr}$, we reach Eq.\
\eqref{eq:KnudsenLimit}. It should be noted, however, that since $l_{\rm
tr}$ depends on the details of the object, {\it e.g.\,}, its shape and the
irregularity of the surface, it is difficult to estimate $l_{\rm tr}$
accurately for complex porous materials like aerogel. We should therefore
regard $l_{\rm tr}$ as a phenomenological parameter comparable to the geometrical
mfp $l_g$.

In order to obtain the expression for the drag force for arbitrary values of Kn, we put
${\bm F} = \bm{F}_0/({1 + A {\rm Kn}})$ and determine the dimensionless
constant $A$ so that the Knudsen-limit formula Eq.\ \eqref{eq:KnudsenLimit} is
recovered. A similar procedure was used by Cunningham \cite{Cunningham1910} in
his theory for the drag force on a sphere immersed in classical fluids, and
the resulting Stokes-Cunningham formula was applied successfully to
Millikan's oil-drop experiment in air \cite{Millikan}. In the present problem, the Cunningham constant
$A$ takes the form of $A = {al_{\rm tr}}/{5\kappa}$. For the sphere model, we
have $A = 6l_{\rm tr}/5l_g$, which illustrates that $A$ is a constant of
order unity.
For the analysis of the experimental results, it is convenient to write
$\bm{F}$ in the following form:
\begin{align}
 \bm{F}
 &= \frac{1}{1 + 1/A{\rm Kn}}\bm{F}_\infty
 = \frac{1}{1 + T^2/AT_1^2}\frac{np_F}{l_{\rm tr}}{\bm v}_{\rm rel}.
\label{DragForce}
\end{align}
Here, $T_1$ is the temperature at which Kn $= 1$. The crossover temperature
$T_1$ is well above $T_a$, as can be seen in Fig.\ \ref{fig:length}.

The Cunningham-type formula implicitly assumes a smooth transition of the
drag force between the hydrodynamic and Knudsen regimes.  Virtanen and
Thuneberg \cite{VirtanenPRL,Virtanen2011} recently reported a detailed
numerical study of the drag force on a vibrating wire immersed in
$^3$He-$^4$He mixtures. Although the system is different from that of
present interest, a common feature can be found in their results, 
showing a smooth transition in the dissipative part of the mechanical impedance
$Z$ at low frequencies.

Having obtained the system of equations, we can now analyze the experiment.
Considering the longitudinal sound of the plane wave form $\propto
e^{i(\bm{q}\cdot\bm{r} -\omega t)}$ and taking account of an experimental fact that
$|c_aq/\omega|^2 \simeq 0.02 \ll 1$, one obtains
\begin{align}
 \alpha &= {\rm Im}\,q = \alpha_{v} + \alpha_{f}, \label{alpha} \\
 \alpha_v &= \frac{2\omega^2p_Fl_a}{15mcc_1^2}\frac{1}{1+T^2/T_a^2}, \label{alphav}\\
 \alpha_f &=\frac{\omega^2ml_{\rm
 tr}}{2cp_F}\frac{(\rho_a/\rho)^2}{1+\rho_a/\rho} \left(1 +
 \frac{T^2}{AT_1^2}\right), \label{alphaf}
\end{align}
where $c = {\omega}/{{\rm Re}\,q} = {c_1}/({1 + \rho_a/\rho})^{1/2}$ and
$c_1$ are the sound velocities in the presence and absence of aerogel,
respectively. The sound velocity $c$ is reduced from $c_1$ by a factor
$(1+\rho/\rho_a)^{-1/2}$ due to the extra inertia from the dragged aerogel.
Using $c_1 = 398$~m/s and $\rho = 0.114$~g/cm$^3$ at 29 bar
\cite{HalperinV1990}, we obtain $c = 338$ m/s, in good agreement with the
experiment.  $c_1$ and therefore $c$ are almost independent of
temperature, which also agrees with our observation. In Eq.\
\eqref{alpha} for sound attenuation, $\alpha_v$ and $\alpha_f$ are associated with the viscous 
damping and the frictional damping caused by the drag force, respectively.  The theoretical calculation represented by the solid lines in Fig.\ \ref{fig:alpha} shows 
excellent agreement with the experiment using the fitting parameters, $l_{\rm tr} $ and $l_{a}$ as listed in the figure caption and $A = 6/5$.
The role of the Cunningham constant $A$ is to shift the crossover temperature
in Eq.~\eqref{alphaf}.
 The best fit to the experimental data
is achieved by taking $A$ to be around $6/5$,
 corresponding to the sphere model with $l_{\rm tr} = l_g$.

To gain insights into the observed behavior, we plot in Fig.\
\ref{fig:alpha} the sound attenuation in the two limiting cases for 29 bar:
the Knudsen ($\alpha_\infty$) and the hydrodynamic ($\alpha_0$) regimes by
taking the limits of $T_1 \to \infty$ and $T_1 \to 0$, respectively in
Eq.~(9). The difference between the two limits comes mainly from the
different $T$-dependence of the frictional damping $\alpha_f$. In the full
Knudsen regime where $T \ll T_{a}$ and effectively $T_1 \to \infty$, both
$\alpha_v$ and $\alpha_f$ are temperature independent.  This simply reflects
the fact that the effective quasiparticle mfp is limited by the geometrical
mfp $l_g$.  The decrease of the Knudsen attenuation above $T_a$ is due to
the decrease of $\alpha_v$. The low temperature experimental data are well
described by the Knudsen behavior. However, as $T$ increases, a significant
deviation occurs especially around $T = T_1$. With increasing $T$ further
above $T_1$, the experimental data approach asymptotically the hydrodynamic
behavior, $\alpha \approx \alpha_{0} \propto T^{2}$.  The crossover
temperature $T_1$ increases with decreasing pressure (see Fig.\
\ref{fig:length}) due to the pressure dependent quasiparticle mean free
path. Consequently, this effect shifts the attenuation minimum to higher
temperature for lower pressure as observed in the experiment.

\begin{acknowledgments}
 This work was supported in part by a Grant-in-Aid for Scientific Research
 (No.\ 21540365) and a Grant-in-Aid for Scientific Research on Innovative
 Areas (No.\ 22103003) from MEXT of Japan, and also by NSF under Grants No. DMR-0803516 (Y.L.) and DMR-0654118 (NHMFL), and the State of Florida.
\end{acknowledgments}

\end{document}